\documentclass[aps,prl,showpacs,twocolumn,showkeys,floats,superscriptaddress,floatfix]{revtex4-1}
\usepackage{bm}
\usepackage{verbatim}
\usepackage{graphicx}
\usepackage{graphics,epsfig}
\usepackage{theorem}
\usepackage{makeidx}
\usepackage{amsmath}
\usepackage{epic}
\usepackage{amscd}
\usepackage{xy}
\usepackage{amssymb}
\usepackage{epsfig}
\begin{document}
\newif\iffigs 
\figstrue
\iffigs \fi
\def\drawing #1 #2 #3 {
\begin{center}
\setlength{\unitlength}{1mm}
\begin{picture}(#1,#2)(0,0)
\put(0,0){\framebox(#1,#2){#3}}
\end{picture}
\end{center} }

\newcommand{\ul}{{\bm u}_{\scriptscriptstyle \mathrm{L}}}
\newcommand{\deltal}{{\delta}_{\scriptscriptstyle \mathrm{L}}}
\newcommand{\ue}{\mathrm{e}}
\newcommand{\ui}{\mathrm{i}\,}
\newcommand{\kg}{{k_{\scriptscriptstyle \mathrm{G}}}}
\def\v{\bm v}
\def\x{\bm x}
\def\k{\bm k}
\def\ds{\displaystyle}

\title{Multiscaling in Hall-Magnetohydrodynamic Turbulence: Insights from a
Shell Model}
\author{Debarghya Banerjee}
\email{debarghya@physics.iisc.ernet.in}
\affiliation{Centre for Condensed Matter Theory,
Department of Physics, Indian Institute of Science,
Bangalore 560012, India}
\author{Samriddhi Sankar Ray}
\email{samriddhisankarray@gmail.com}
\altaffiliation[\\ Formerly at ]{Laboratoire Lagrange, OCA, UNS, CNRS, BP 4229,
06304 Nice Cedex 4, France}
\affiliation{International Centre for Theoretical Sciences, Tata Institute of Fundamental Research, Bangalore 560012,
India}
\author{Ganapati Sahoo}
\email{ganapati.sahoo@ds.mpg.de}
\affiliation{Max Planck Institute for Dynamics and Self-Organization,
Am Fassberg 17, 37077 G\"ottingen, Germany}
\author{Rahul Pandit}
\email{rahul@physics.iisc.ernet.in}
\altaffiliation[\\ Also at ]{Jawaharlal Nehru Centre For Advanced
Scientific Research, Jakkur, Bangalore, India}
\affiliation{Centre for Condensed Matter Theory,
Department of Physics, Indian Institute of Science,
Bangalore 560012, India}

\begin{abstract}

We show that a shell-model version of the three-dimensional
Hall-magnetohydrodynamic (3D Hall-MHD) equations provides a natural theoretical
model for investigating the multiscaling behaviors of velocity and magnetic
structure functions. We carry out extensive numerical
studies of this shell model, obtain the scaling exponents 
for its structure functions, in both the
low-$k$ and high-$k$ power-law ranges of 3D Hall-MHD, and find that the
extended-self-similarity (ESS) procedure is helpful in extracting the multiscaling 
nature of structure functions in the high-$k$ regime, which otherwise appears
to display simple scaling. Our results shed light on intriguing solar-wind 
measurements.

\end{abstract}


\date{\today}

\maketitle

Turbulent plasmas abound in accretion disks, galaxies, stars, the solar wind,
and laboratory experiments~\cite{mhd,cowley07}; thus, the characterization of
the statistical properties~\cite{mhd,sahoo11,mhddns} of turbulence in such
plasmas is a problem of central importance in astrophysics, plasma physics,
fluid dynamics, and nonequilibrium statistical mechanics. Such a
characterization begins with the energy spectra: e.g., in homogeneous and
isotropic fluid turbulence the energy spectrum $E(k)$, which gives the
distribution of energy over different wave numbers $k$, assumes the scaling
form $E(k) \sim k^{-\alpha}$ if the Reynolds numbers $Re$ is large and $k$ is
in the inertial range $L^{-1} \ll k \ll k_d$, where $L$ is the energy-injection
length scale and $k_d \equiv 2 \pi / \eta_d$, with $\eta_d$ the length scale at
which viscous dissipation becomes significant; the phenomenological theory of
Kolmogorov (K41) yields~\cite{K41,frisch95} $\alpha = 5/3$. Turbulent plasmas
show similar scaling forms for the kinetic and magnetic-energy spectra $E^u(k)$
and $E^b(k)$, if the turbulence is statistically homogeneous and isotropic, and
both $Re$ and the magnetic Reynolds numbers $Re_M$ are large; their ratio $Pr_M
= Re_M/Re$, the magnetic Prandtl number, governs the relative sizes of the
fluid and magnetic dissipation length scales $\eta_d^u$ and $\eta_d^b$; the
inertial-range scaling properties of $E^u(k)$ and $E^b(k)$ have been studied
theoretically and numerically by using the equations of magnetohydrodynamics
(MHD)~\cite{mhd,sahoo11,mhddns}.  Energy-spectra measurements in the solar wind
~\cite{solarwind} have shown, however, that $E^b(k)$ displays \textit{two}
power-law ranges. Several
authors~\cite{vkrishan,shaikh09,meyrand12,hallmhdshell,galtiershell} have
suggested that, to obtain these two power-law regimes, we must augment the MHD
equations with a Hall-effect term, which leads to a scale separation at the
ion-inertial length $d_I$ or, equivalently, at the wave number $k_I = 2\pi/d_I$. For $k < k_I$,
$E^b(k) \propto k^{-\alpha^{b,1}}$, it has been observed that $\alpha^{b,1}
\simeq 5/3$. For $k_d > k > k_I$, $E^b(k) \propto k^{-\alpha^{b,2}}$,
where $\alpha^{b,2}$ is either $\simeq7/3$ or $\simeq11/3$. 
The value of $\alpha^{b,2}$ 
depends on whether the magnetic energy dominates over the fluid kinetic energy,
which occurs in the electron-MHD (EMHD)~\cite{biskamp} limit, or the converse, i.e.,
the ion-MHD (IMHD) limit. These limits follow from the 3D Hall-MHD equations:
EMHD is obtained if the induction term is sub-dominant to the Hall term; 
in the IMHD case these two terms are comparable to each other. In the EMHD limit,
we obtain a single, characteristic scale and K41 phenomenology yields $\alpha^{b,2} = 7/3$;
in the IMHD case a comparison of the transfer time, from the Hall-term, and a second
time, from the induction part, followed by simple dimensional analysis yields 
$\alpha^{b,2} = 11/3$~\cite{galtiershell,meyrand12}.

Direct numerical simulations (DNSs)~\cite{shaikh09,meyrand12,hallmhdshell} have
just begun to resolve these two scaling ranges; but their spatial resolution is
much more limited than has been achieved in DNS studies of MHD
turbulence~\cite{sahoo11,mhddns}. Thus, they have not been used to study the
scaling or multiscaling properties of order $p$ fluid and magnetic structure functions
(defined below). However, measurements of such equal-time magnetic structure functions 
in solar-wind measurements~\cite{solarwind} show
that, although there is significant multiscaling in the low-$k$ ($k < k_I$),
power-law range of $E^b(k)$, the scaling exponents in the second, high-$k$
($k_d > k > k_I$) power-law range increase linearly with the order $p$. Thus,
it behooves us to develop a theoretical understanding of these important and
intriguing observations and to test them.

\begin{table*}
\label{table:param}
\begin{center}
\begin{tabular}{|l|c|c|c|c|c|c|c|c|}
\hline
Runs & $Pr_M$ & $\nu$ & $\nu_2$ & $\eta_2$ & $d_I$ & ${Pr_M}_{\rm{eff}}$ & $\nu_{\rm{eff}}(\times 10^{-6})$ & $\eta_{\rm{eff}}(\times 10^{-6})$ \\ \hline
R1 & $1$ & $10^{-8}$ & $5 \times 10^{-13}$ & $5 \times 10^{-13}$ & $0.1$ & $1.2 \pm 0.3$ & $1.8 \pm 0.2$ & $1.6 \pm 0.4$         \\ \hline
R2 & $10$ & $10^{-7}$ & $5 \times 10^{-12}$ & $5 \times 10^{-13}$ & $0.1$ & $4.3 \pm 0.8$ & $4.7 \pm 0.4$ & $1.1 \pm 0.2$    \\ \hline
R3 & $10$ & $10^{-8}$ & $1 \times 10^{-13}$ & $1 \times 10^{-14}$ & $0.01$ & $2.8 \pm 0.6$ & $0.74 \pm 0.09$ & $0.27 \pm 0.06$        \\ \hline
R4 & $1$ & $10^{-9}$ & $1 \times 10^{-14}$ & $1 \times 10^{-14}$ & $0.01$ & $1.0 \pm 0.2$ & $0.31 \pm 0.04$ & $0.31 \pm 0.07$     \\ \hline
\hline
Runs & $\tau$ & $l_{I}$ & $Re_{\lambda}(\times 10^{5})$ & $\eta_d^u(\times 10^{-4})$ & $\eta_d^b(\times 10^{-4})$ & $\alpha^u$ &$\alpha^{b,1}$ &$\alpha^{b,2}$\\  \hline
R1 & $16.3 \pm 0.6$ & $6.64 \pm 0.09$ & $11 \pm 2$ & $1.69 \pm 0.04$ & $2.2 \pm 0.3$ & $1.66 \pm 0.03$ & $1.75 \pm 0.03$ & $3.39 \pm 0.08$       \\ \hline
R2 & $16 \pm 1$ & $6.7 \pm 0.1$ & $4.9 \pm 0.8$ & $3.40 \pm  0.08$ & $1.7 \pm 0.2$ & $1.65 \pm 0.03$ & $1.69 \pm 0.03$ & $3.45 \pm 0.06$        \\ \hline
R3 & $16 \pm 1$ & $6.6 \pm 0.1$ & $26 \pm 4$ & $0.90 \pm 0.03$ & $0.62 \pm 0.07$ & $1.69 \pm 0.03$ & $1.74 \pm 0.02$ & $3.34 \pm 0.09$      \\ \hline
R4 & $13.3 \pm 0.7$ & $6.8 \pm 0.1$ & $64 \pm 9$ & $0.46 \pm 0.02$ & $0.64 \pm 0.07$ & $1.70 \pm 0.02$ & $1.74 \pm 0.02$ & $3.28 \pm 0.08$      \\ \hline
\end{tabular}
\end{center}
\caption{The values of the different parameters (see text) used in our runs ${\rm R1-R4}$
and the spectral exponents  $\alpha^u$, $\alpha^{b,1}$, and $\alpha^{b,2}$ 
obtained.}
\end{table*}

We show that a shell-model version of the 3D Hall-MHD
equations~\cite{hallmhdshell,galtiershell}, which is a generalization of MHD shell
models~\cite{basu,mhdshell}, provides a natural theoretical model for
investigating such multiscaling behaviors in structure functions in 3D Hall-MHD
turbulence. Given the large range of scales that we can cover in this shell
model~\cite{goy}, its magnetic spectrum $E^b(k)$ reveals two, distinct, power-law ranges. 
We carry out the most comprehensive numerical study of this 3D Hall-MHD shell
model attempted so far; and thereby we characterize and quantify, 
for the first time, the properties of the order-$p$ magnetic and 
velocity structure functions in this model via their scaling exponents 
$\zeta_p^u$ (fluid), $\zeta_p^{b,1}$ (magnetic, $k < k_I$ regime), and
$\zeta_p^{b,2}$ (magnetic, $k_d > k > k_I$ regime). We find that all 
three sets of exponents show clear signatures of multiscaling. In particular, 
we find the remarkable result that magnetic structure functions display 
multiscaling for {\it both} the low-$k$ {\it and} the high-$k$ 
power-law ranges. 
A second significant and surprising 
finding is that, although the exponents $\zeta_p^{b,2} \ne \zeta_p^{b,1}$,
$\zeta_p^{b,2} \ne \zeta_p^u$, 
the exponent ratios 
$\frac{\zeta_p^{b,2}}{\zeta_3^{b,2}} \simeq \frac{\zeta_p^{b,1}}{\zeta_3^{b,1}} \simeq \frac{\zeta_p^u}{\zeta_3^u}$
~\cite{foot1}. 

\begin{figure*}
\includegraphics[width=5.9cm,height=4cm]{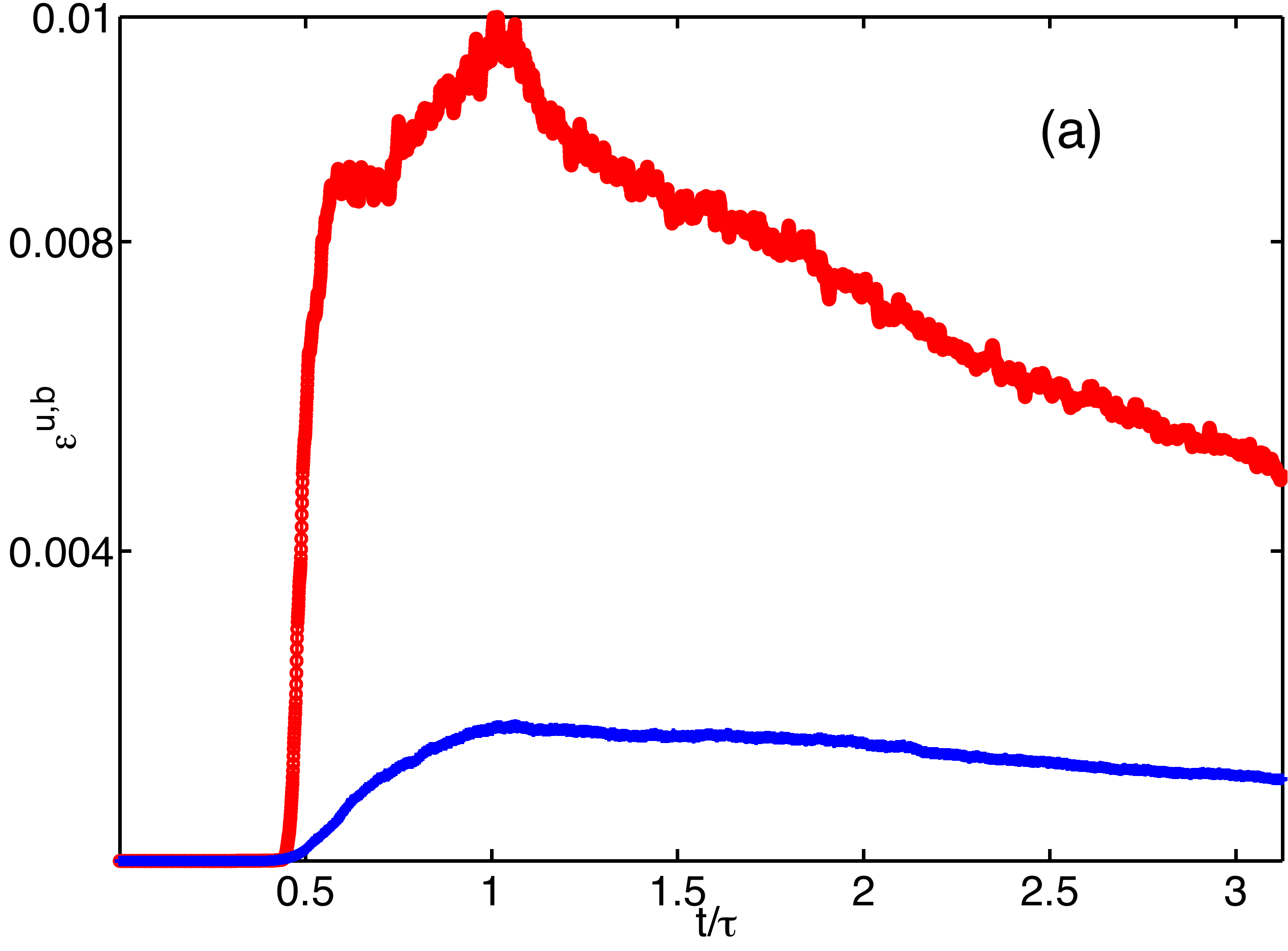}
\includegraphics[width=5.9cm,height=4cm]{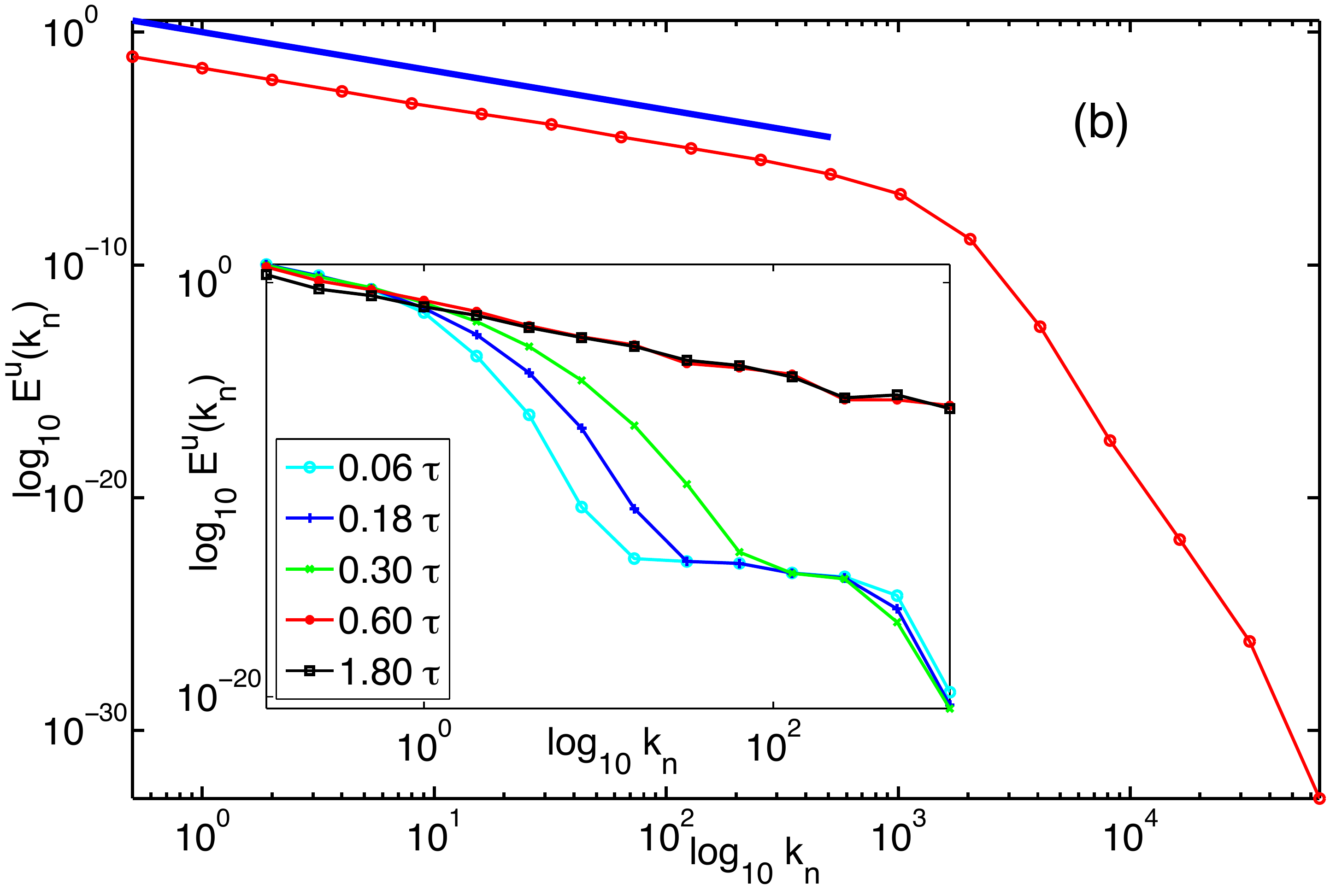}
\includegraphics[width=5.9cm,height=4cm]{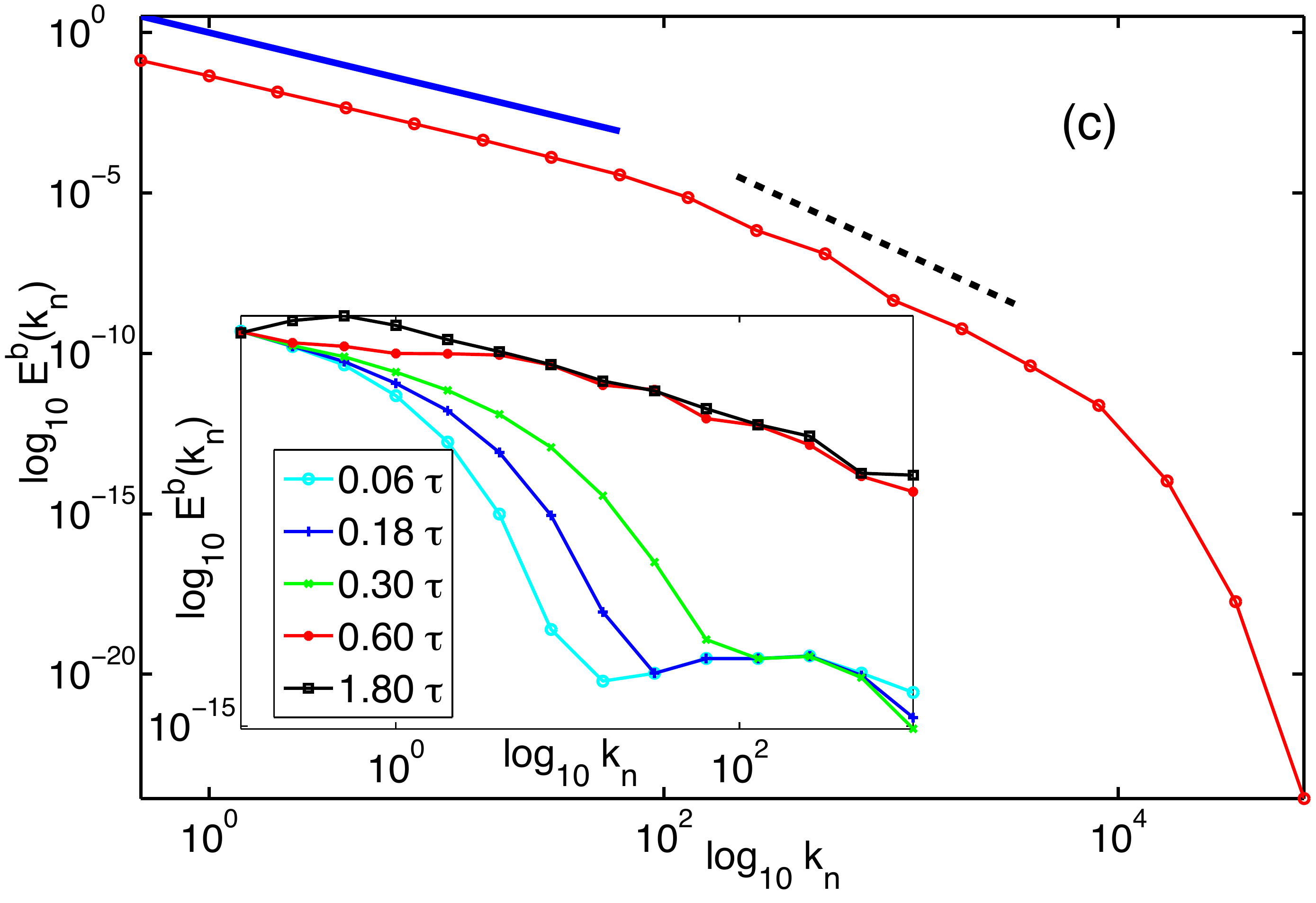}
\caption{(Color online) (a) Plots of  $\varepsilon^{u}$ (red, upper 
curve) and  $\varepsilon^{b}$ (blue, lower curve) versus the rescaled time $t/\tau$.
(b) Plot of the kinetic-energy spectrum $E^u(k)$; the thick, blue line indicates the K41-scaling power law. 
Inset : Time evolution of  
$E^u(k)$ during the initial period with transients; the legend indicates the time at which each curve is obtained.
(c) Plot of the magnetic-energy spectrum $E^b(k)$; the thick, blue line indicates K41 scaling in the low-$k$ region and 
the dashed, black line indicates $k^{-3.5}$ scaling in the high-$k$ region . Inset : Time evolution of  
$E^b(k)$ during the initial period with transients; the legend indicates the time at which each curve is obtained.
Here we use $E^u(k_n)=\Sigma^u_2(k_n)/k_n$ and $E^b(k_n)=\Sigma^b_2(k_n)/k_n$ to suppress 3-cycles (see text).
The data shown are from run R2 and the spectral exponents for all our runs are reported 
in Table I similar plots for runs R1, R3, and R4 are given in the Supplemental Material~\cite{supple}.}
\label{figs:spectra}
\end{figure*}

The 3D Hall-MHD equations for the velocity $\bf{u}$ and magnetic
$\bf{b}$ fields are
\begin{eqnarray}
\label{eq:hallmhd}
\frac{\partial \bf{u}}{\partial t} + (\bf{u} \cdot \nabla) \bf{u} & = &-\nabla p + \bf{j} \times \bf{b} + \nu \nabla^{2} \bf{u} ;  \nonumber \\
\frac{\partial \bf{b}}{\partial t} & = & \nabla \times [({\bf u} - d_I {\bf j}) \times {\bf b}] + \eta \nabla^{2} \bf{b}; 
\end{eqnarray}
here $\nu$ and $\eta$ are the kinematic viscosity and
magnetic diffusivity, respectively, $d_I$ is the ion-inertial 
length, the scale at which the Hall effect
becomes important, the current density vector $\bf{j} =
\nabla \times \bf{b}$, the pressure is $p$, $\nabla \cdot
\bf{b} = 0$, and, at low Mach numbers, the flow is
incompressible, i.e., $\nabla \cdot \bf{u} = 0$. We
define the dissipation length scales  $\eta_d^u =
(\nu^3/\varepsilon^u)^{1/4}$ and $\eta_d^b =
(\eta^3/\varepsilon^b)^{1/4}$, where $\varepsilon^u$ and
$\varepsilon^b$ are the kinetic and magnetic-energy
dissipation rates, respectively; we restrict ourselves to
decaying turbulence, so we do not include forcing terms.
The Hall term, which is a singular perturbation of the
MHD equations~\cite{meyrand12}, has a significant effect
if $d_I \gg \eta_d^u,\,\eta_d^b$. The shell-model
versions of Eq.(\ref{eq:hallmhd}) are~\cite{hallmhdshell,galtiershell}:
\begin{eqnarray}
\label{Eq:shell}
\frac{du_{n}}{dt} & = & - \nu k_{n}^{2}u_{n} - \nu_2
k_{n}^{4}u_{n} + \iota [\Phi_n^u]^{*} ,
\nonumber \\
\frac{db_{n}}{dt} & = & - \eta k_{n}^{2}b_{n} - \eta_2
k_{n}^{4}b_{n} + \iota[\Phi_n^b]^{*},
\end{eqnarray}
where $u_n$ and $b_n$ are, respectively, the complex velocity and magnetic
field in the shell $n$, $*$ denotes complex conjugation, $1 \leq n \leq N$,
where $N$ is the total number of shells, $\Phi_n^u = A_{n}(u_{n+1}u_{n+2} -
b_{n+1}b_{n+2}) + B_{n}(u_{n-1}u_{n+1} - b_{n-1}b_{n+1}) + C_{n}(u_{n-2}u_{n-1}
- b_{n-2}b_{n-1})$ and $\Phi_n^b = D_{n}(u_{n+1}b_{n+2} - b_{n+1}u_{n+2}) +
  E_{n}(u_{n-1}b_{n+1} - b_{n-1}u_{n+1}) - F_{n}(u_{n-2}b_{n-1} -
b_{n-2}u_{n-1})
- d_I[G_{n}b_{n+1}b_{n+2} + H_{n}b_{n-1}b_{n+1} + I_{n}b_{n-2}b_{n-1}]$, with
  $A_{n}=k_{n}$, $B_{n} = -\frac{1}{2}k_{n-1}$, $ C_{n} = -\frac{1}{2}k_{n-2}$,
$D_{n}=\frac{1}{6}k_{n}$, $ E_{n} = \frac{1}{3}k_{n-1}$, $ F_{n} = 
\frac{2}{3}k_{n-2}$, $G_{n}=-\frac{1}{2}(-1)^{n}k_{n}^{2}$, $ H_{n} = -
\frac{1}{2} (-1)^{n-1}k_{n-1}^{2}$, $I_{n}= (-1)^{n-2}k_{n-2}^{2}$, $k_n = 2^n
k_0$, and $k_0 = 1/16$; the values of the coefficients $A_n - I_n$ are
determined by enforcing the shell-model analogs of the Hall-MHD conservation
laws, in the inviscid, unforced limit; the conserved quantities are the total
energy $E = \Sigma_n (|u_n|^2 + |b_n|^2)/2$, the magnetic helicity $H_M =
\Sigma_n (-1)^n|b_n|^2/2k_n$, and the ion helicity $H_I = 
\Sigma_n\left ((b_nu_n^* + b_n^*u_n) + d_I (-1)^nk_n|u_n|^2/2\right )$; the hyperviscosity
$\nu_2$ and the magnetic hyperdiffusivity $\eta_2$ have to be included for
numerical stability~\cite{meyrand12,galtiershell}. We use the boundary
conditions $A_{N-1}=A_N = B_1 = B_N = C_1 = C_2 = 0$, $D_{N-1}=D_{N} = E_1 =
E_N = F_1 = F_2 = 0$, $G_{N-1}=G_{N} = H_1 = H_N = I_1 = I_2 = 0,$ and the
following initial values for $u_n = u^{(0)} k_{n}^{-1/3}e^{-k_{n}^{2}+\iota
\phi^u}$ and $b_{n} = b^{(0)} k_{n}^{-1/3}e^{-k_{n}^{2} + \iota \phi^b}$; here
$u^{(0)} = 0.5$, $ b^{(0)} = 0.05$, and the random phases $\phi^u$ and $\phi^b$
are distributed uniformly on the interval $ [-\pi,+\pi]$; different values of
these random phases distinguish different initial conditions; we work with decaying turbulence, 
so there is no forcing term; and our results are
averaged over $7500$ independent {\it initial} conditions ~\cite{foot2}. We set $N=22$, use a
second-order, slaved Adams-Bashforth scheme~\cite{cox} for solving the shell-model
ordinary differential equations (\ref{Eq:shell}), and calculate the energy spectra
$E^u(k_n) = \frac{1}{2}|u_n|^2/k_n$ and $E^b(k_n) = \frac{1}{2}|b_n|^2/k_n$
(the superscripts $u$ and $b$ refer to velocity
and magnetic field, respectively), the root-mean-square velocity
$u_{rms}=\sqrt{\Sigma_n \mid u_n \mid^2}$, the Taylor microscale $\lambda =
\sqrt{\Sigma_n E^u(k_n)/\Sigma_n k_n^2 E^u(k_n)}$, the Taylor-microscale
Reynolds number $Re_{\lambda}=u_{rms} \lambda /\nu_{\rm{eff}}$, the integral
length scale $l_I=\Sigma_n \left ( E^u(k_n)/k_n \right )/\Sigma_n E^u(k_n)$, the effective
viscosity and magnetic diffusivity $\nu_{\rm{eff}} = \Sigma_n \left (\nu k_n^2
E^u(k_n) + \nu_2 k_n^4 E^u(k_n) \right )/\Sigma_n k_n^2 E^u(k_n)$ and $\eta_{\rm{eff}} =
\Sigma_n \left (\eta k_n^2 E^b(k_n) + \eta_2 k_n^4 E^b(k_n)\right )/\Sigma_n k_n^2 E^b(k_n)$,
respectively, the effective magnetic Prandtl number~\cite{iskakov}
${Pr_M}_{\rm{eff}}=\nu_{\rm{eff}}/\eta_{\rm{eff}}$, and the dissipation rates 
$\varepsilon^u = \nu_{\rm{eff}}\Sigma_n k_n^2 E^u(k_n)$ and $\varepsilon^b = \eta_{{\rm
eff}}\Sigma_n k_n^2 E^b(k_n)$. The parameters of our simulations are given in Table I.

In shell models, the equal-time, order-$p$ structure functions for the velocity field and the magnetic 
field are defined, respectively, as $S^u_p(k_n) = \langle |u_n|^p \rangle \sim k_n^{\zeta^u_p}$ and  
$S^b_p(k_n) = \langle |b_n|^p \rangle$, where $S^b_p(k_n) \sim k_n^{\zeta_p^{b,1}}$ ($k < k_I$) and 
$S^b_p(k_n) \sim k_n^{\zeta_p^{b,2}}$ ($k_d > k > k_I$).
However, to remove the effects of an underlying three
cycle in GOY-type shell models~\cite{basu,goy}, we use the modified structure functions
$\Sigma_p^u(k_n) = \langle | \Im[u_{n+2}u_{n+1}u_{n} +
1/4u_{n-1}u_{n}u_{n+1}]|^{p/3} \rangle$ and $\Sigma_p^b(k_n) = \langle |
\Im[b_{n+2}b_{n+1}b_{n} + 1/4b_{n-1}b_{n}b_{n+1}]|^{p/3} \rangle,$ from which
we can obtain multiscaling exponents via $\Sigma_p^u(k_n) \sim
k_n^{\zeta_p^u}$, $\Sigma_p^b(k_n) \sim k_n^{\zeta_p^{b, 1}}$ ($k_n < k_I$), and 
$\Sigma_p^b(k_n) \sim k_n^{\zeta_p^{b, 2}}$ ($k_d > k_n > k_I$).
We also use the extended self-similarity (ESS) procedure~\cite{ess} to determine
exponent ratios from slopes of log-log plots of $\Sigma_{p}^{u}$ versus $\Sigma_{3}^{u}$
and their magnetic counterparts (Fig. 2 inset).

\begin{figure}
\includegraphics[width=1.0\columnwidth]{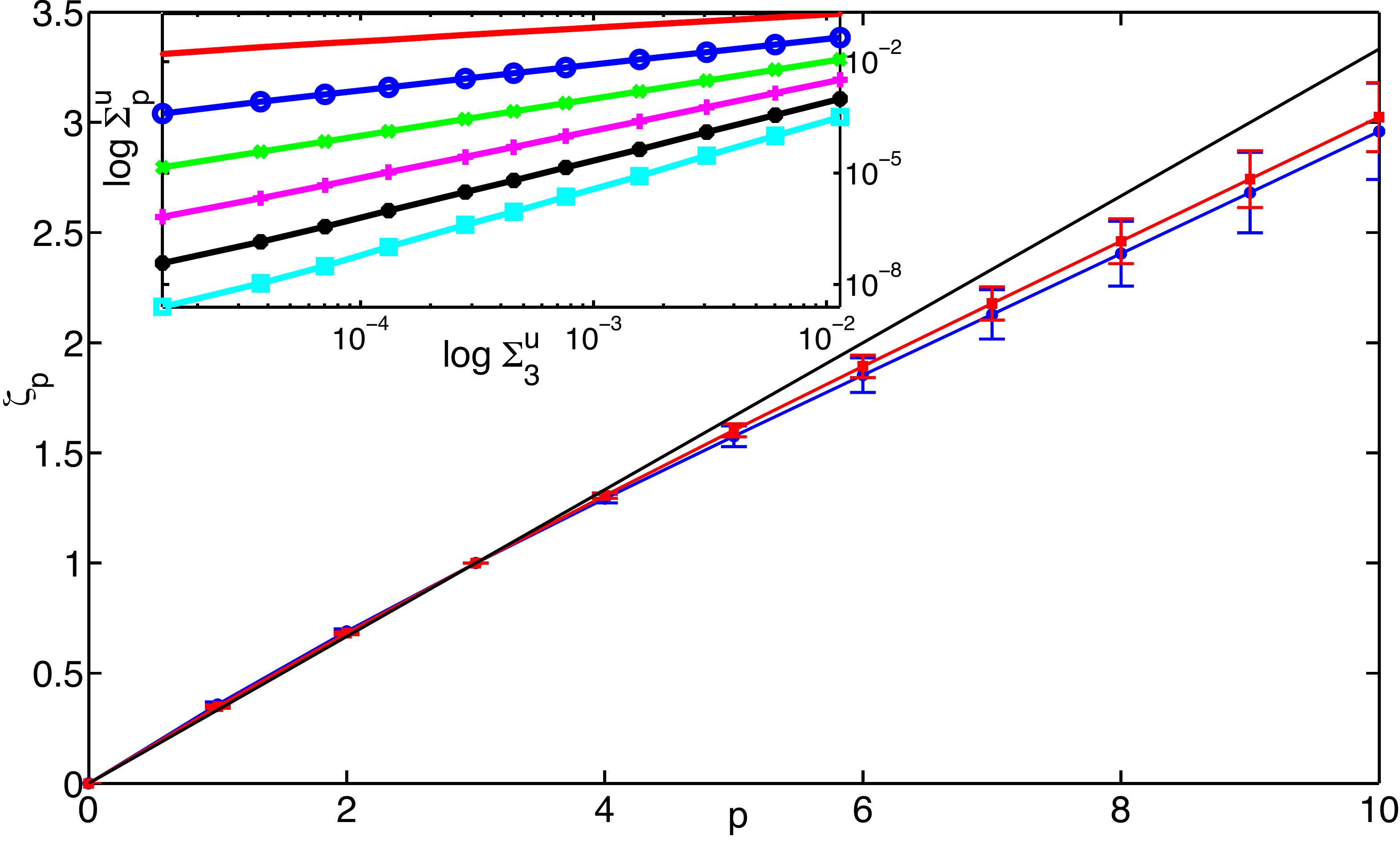}
\caption{(Color online) The exponents $\zeta_p^u$ (blue filled circles connected by a blue line),  
$\zeta_p^{b,1}$ (red filled squares connected by a red line), and the K41 prediction (thick black line) versus $p$.
The lines connecting the data points are a guide to the eye.
(Inset) A representative plot of $\Sigma_p^u$ versus $\Sigma_3^u$; equal-time 
exponents are calculated from such plots. The different curves are for $p = 1 \,{\rm (top)} \ldots 6$ (bottom) 
and are from run R2.} 
\label{figs:zetap_fl_mg}
\end{figure}

In Fig. 1(a) we show plots of $\varepsilon^{u}$ (red, upper curve) and
$\varepsilon^{b}$ (blue, lower curve) versus the rescaled time $t/\tau$, for
run R2, where the box-size eddy-turnover time $\tau = 1/(u_1 k_1)$ is evaluated
at the principal peak of $\varepsilon^{u}$.  This peak signals the completion
of the Richardson cascade~\cite{sahoo11}, as we can see from the time evolution
of $E^u(k_n)$ and $E^b(k_n)$, in the insets of Figs.1(b) and (c), respectively,
where the red lines with full circles denote the spectra at cascade completion.
We evaluate the spectral-slope exponent $\alpha^u$ 
at cascade completion from log-log plots of $E^u(k_n)$ versus $k_n$ as
shown in Fig. 1(b). We find $\alpha^u \simeq 5/3$, as predicted by dimensional analysis, 
and illustrated in Fig. 1(b) by a thick, blue line.
In Fig. 1(c) we show a representative plot of $E^b(k_n)$; we 
see two different scaling regimes clearly: (1) from the low-$k$ one  
(solid, blue line), we find that $\alpha^{b,1} \simeq 5/3$, which is consistent with dimensional analysis;
(2) from the high-$k$ regime, we obtain $\alpha^{b,2} = 3.45 \pm 0.06$, which is close to the 
dimensional-analysis value 11/3 for IMHD systems, such as ours, in which induction and 
Hall terms are comparable~\cite{galtiershell,meyrand12}. These spectral exponents are 
consistent with those in solar-wind experiments~\cite{solarwind}. 
In Table I we provide our results for all three spectral exponents;
we obtain the values of these and all other exponents from the means of our runs with 7500
independent initial conditions; the error bars follow from the associated
standard deviations. 

To characterize the statistical properties of the Hall-MHD system, we now calculate the 
equal-time exponents $\zeta_p^u$, $\zeta_p^{b,1}$ ($k < k_I$), and $\zeta_p^{b,2}$ ($k_d > k > k_I$) 
via the modified structure functions $\Sigma_p^u$ and $\Sigma_p^b$, just 
after cascade completion. We find $\zeta_3^u = 1 = \zeta_3^{b,1}$, which is consistent with dimensional analysis. 
In Fig. 2 we show the order-$p$ equal-time exponents $\zeta_p^u$ (blue, filled 
circles) and $\zeta_p^{b,1}$ (red, filled squares) for integer values of $p$ between 1 and 10; 
the thick, black line illustrates the dimensional, simple K41 scaling.   
We see that $\zeta_p^u \simeq \zeta_p^{b,1}$ and both exponents show clear multiscaling corrections
to K41 scaling, with values  
consistent with those obtained in 3D MHD turbulence~\cite{sahoo11}. 
We obtain these multiscaling exponents by using $\Sigma_p^u$ and $\Sigma_p^b$ 
and the ESS procedure~\cite{ess},  
to extend the scaling range. However, the result $\zeta_3^u \simeq 1$ ensures that 
the exponent ratios and the exponents themselves are equal (within error bars).  
In Table II, we list the order-$p$ equal-time exponents $\zeta_p^u$ 
and $\zeta_p^{b,1}$ for integer values of $p$ between 1 and 10.

\begin{table}
\label{table:multi}
\begin{center}
\begin{tabular}{|c|c|c|c|c|}
\hline
$p$ &  $\zeta_{p}^{u}$  &  $\zeta_{p}^{b,1}$  &  $\zeta_{p}^{b,2}$ &  $\zeta_{p}^{b,2}/\zeta_{3}^{b,2}$ 
 \\ \hline
$1$  &  $0.36 \pm 0.01$  &  $0.348 \pm 0.006$  &  $1.40 \pm 0.04$    & $0.38 \pm 0.01$   \\ 
$2$  &  $0.69 \pm 0.01$  &  $0.682 \pm 0.006$  &  $2.6 \pm 0.1$      & $0.698 \pm 0.009$    \\
$3$  &  1.0                &  1.0              &  $3.7 \pm 0.2$      & 1.0               \\
$4$  &  $1.29 \pm 0.02$  &  $1.31 \pm 0.01$    &  $4.8 \pm 0.3$      & $1.30 \pm 0.01$   \\
$5$  &  $1.57 \pm 0.06$  &  $1.60 \pm 0.03$    &  $6.0 \pm 0.4$      & $1.60 \pm 0.03$   \\
$6$  &  $1.85 \pm 0.09$  &  $1.89 \pm 0.05$    &  $7.1 \pm 0.4$      & $1.90 \pm 0.04$   \\
$7$  &  $2.1 \pm 0.1$    &  $2.18 \pm 0.08$    &  $8.2 \pm 0.6$      & $2.20 \pm 0.06$   \\
$8$  &  $2.4 \pm 0.2$    &  $2.5 \pm 0.1$      &  $9.3 \pm 0.6$      & $2.50 \pm 0.07$   \\
$9$  &  $2.7 \pm 0.2$    &  $2.7 \pm 0.1$      &  $10.5 \pm 0.7$     & $2.81 \pm 0.08$   \\
$10$ &  $3.0 \pm 0.3$    &  $3.0 \pm 0.2$      &  $11.6 \pm 0.8$     & $3.11 \pm 0.09$   \\
\hline
\end{tabular}
\end{center}
\caption{Multiscaling exponents $\zeta_{p}^{u}$, $\zeta_{p}^{b,1}$, $\zeta_{p}^{b,2}$, and the exponent ratio $\zeta_{p}^{b,2}/\zeta_{3}^{b,2}$ for 
run R2. The exponents from the other runs are equal to the ones shown here. We note that $\zeta_{p}^{u} \simeq \zeta_{p}^{b,1} \simeq \zeta_{p}^{b,2}/\zeta_{3}^{b,2} \ne \zeta_{p}^{b,2}$~\cite{supple}.}
\end{table}

We finally turn to the exponents $\zeta_p^{b,2}$, which characterize the
high-$k$ regime ($k_d > k > k_I$). Solar-wind measurements~\cite{solarwind} of
$\zeta_p^{b,2}$ suggest simple-scaling behaviour, with $\zeta_p^{b,2}$
a linear function of $p$. In Fig. 3 we plot $\zeta_p^{b,2}$
(obtained without ESS) versus $p$. Our results for $\zeta^{b,2}_p$ are in
qualitative agreement with solar-wind measurements to the extent that there is
only mild multiscaling; i.e., $\zeta^{b,2}_p$ is a {\it nonlinear}, monotone, 
increasing function of $p$, but the deviation from a linear dependence on $p$ 
is not very pronounced. 
Although the exponents $\zeta^{b,2}_p$, for all the runs 
R1-R4, are in agreement with each other, given our error-bars, their mean values
seem to decrease with $d_I$. 
We now use the ESS procedure to obtain the exponent ratios $\zeta_p^{b,2}/\zeta_3^{b,2}$, which
are plotted versus $p$ in the inset of Fig. 3 (note $\zeta_3^{b,2} \ne 1$). This ESS plot is remarkable for two reasons~: (1)
There is a clear signature of multiscaling (the thick, black line in the inset indicates simple scaling);
(2) although the exponents $\zeta_p^{b,2}$ are very different from $\zeta_p^u$ and
$\zeta_p^{b,1}$, the ratios $\zeta_p^{b,2}/\zeta_3^{b,2}$ are equal to
$\zeta_p^u$ and $\zeta_p^{b,1}$ (within error bars). In Table II, we list both $\zeta_p^{b,2}$ and
$\zeta_p^{b,2}/\zeta_3^{b,2}$ for different values of $p$ for the representative run R2.

\begin{figure}
\includegraphics[width=1.0\columnwidth]{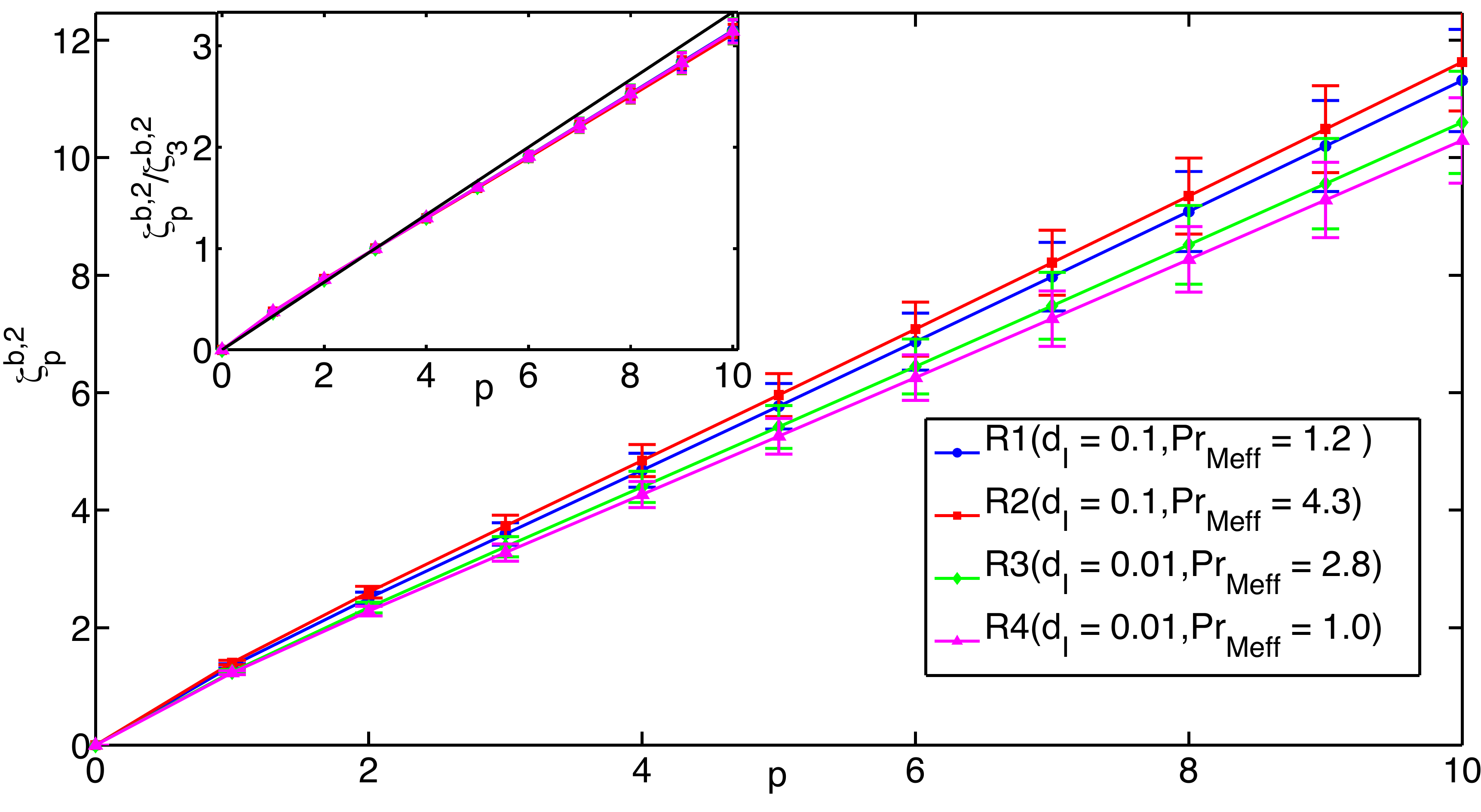}
\caption{(Color online) Plot of the exponents $\zeta_p^{b,2}$ versus the order $p$ showing  
a mild departure from linear scaling. Inset : the exponent ratios  $\zeta_p^{b,2}/\zeta_3^{b,2}$ (obtained via ESS)  
versus $p$, showing clear multiscaling. The thick, black line indicates simple scaling.
The Exponents $\zeta_p^{b,2}$  and $\zeta_p^{b,2}/\zeta_3^{b,2}$ are shown for runs R1-R4 (see legend).} 
\label{figs:zetap_mg_hall}
\end{figure}
 
We hope our extensive studies of the multiscaling of structure functions
in a shell model for 3D Hall-MHD will stimulate high-precision and high-resolution 
experimental and DNS
studies to determine conclusively whether 3D Hall-MHD turbulence shows 
multiscaling for $k_d > k > k_I$. Our ESS results suggest  
that structure functions show mild, but distinct, multiscaling in this region
To obtain quantitative agreement with solar-wind  
exponents, we must, of course, carry out DNS studies of the 3D Hall-MHD
equations (\ref{eq:hallmhd}) and include compressibility effects and a 
mean magnetic field~\cite{vkrishan,shaikh09,meyrand12}; however, current 
computational resources limit severely the spatial resolution of 
such DNS studies so they cannot (a) uncover the multiscaling of magnetic-field
structure functions in 3D Hall-MHD turbulence in both low- and high-$k$
power-law ranges and (b) obtain well-averaged multiscaling exponent ratios.
For the moment, therefore, the shell-model study, which we have undertaken, 
provides the only way of understanding the multiscaling of structure functions
in the solar wind~\cite{solarwind} and the apparent and intriguing universality of 
the exponent ratios.  
This apparant universality needs to be investigated in detail in experiments and DNS.

\begin{acknowledgements}
We thank A. Basu and V. Krishan for discussions, S. Galtier for the preprint of
Ref.~\cite{meyrand12}, CSIR, UGC, and DST (India) for support, and SERC (IISc)
for computational resources.  R.P. and G.S. are members of the International
Collaboration for Turbulence Research; G.S., R.P., and S.S.R. acknowledge support from
the COST Action MP0806; S.S.R. thanks the European Research Council for support under the European 
Community’s Seventh Framework Program (FP7/2007-2013, Grant Agreement no. 240579).
\end{acknowledgements}



\appendix

\section{Supplemental Material}

In this Supplemental Material we describe details of our work that are of
interest only to specialists of  the field. The notations and abbreviations
used in this Supplemental Material are the same as in the main paper.

In our main paper, we discuss the statistical nature of turbulence in the
Hall-MHD plasma and, in particular, the scaling properties of various structure
functions, in great detail.  To substantiate our claims, we show in the main
paper representative data from only a single set of simulations (except in Fig.
3 where we show results from all our simulations), namely, Run R2 (see Table I
of the main paper). In Fig. 3 of the main paper we do show exponents from all
the four different sets of simulation (as detailed in Table I of the main
paper); however, in Table II (main paper) we list exponents from R2 only; because
the different sets of simulations all agree with each other, representative
data from one set of simulations (R2), in the main paper, is enough to
highlight the nature of multiscaling in Hall-MHD turbulence.

We give here the equal-time exponents from the runs
R1, R3, R4 in Table I; these exponents are in agreement with the ones listed for
R2 in Table II of the main paper.  Furthermore, we show plots for the fluid and
magnetic energy dissipation rates (Fig 1(a) for run R1, Fig 2(a) for run R3, and Fig
3(a) for run R4), the kinetic energy spectrum (Fig 1(b) for run R1, Fig 2(b) for run R3,
and Fig 3(b) for run R4), and the magnetic energy spectrum (Fig 1(c) for run R1, Fig
2(c) for run R3, and Fig 3(c) for run R4); these plots are analogous to the plots shown
in Fig.1, for run R2, in the main paper.

The parameters of the runs R1, R3, and R4 (along with those for run R2) are given in
Table (I) of our main paper.

\begin{table*}[HTBP]
\label{table:multi2}
\begin{center}
\begin{tabular}{||l|c|c|c||c|c|c||c|c|c||}
\hline &
\multicolumn{3}{c||}{R1} & \multicolumn{3}{c||}{R3} & \multicolumn{3}{c||}{R4}\\
\hline
$p$ &  $\zeta_{p}^{u}/\zeta_{3}^{u}$  &  $\zeta_{p}^{b,1}/\zeta_{3}^{b,1}$  &  $\zeta_{p}^{b,2}/\zeta_{3}^{b,2}$ &  $\zeta_{p}^{u}/\zeta_{3}^{u}$  &  $\zeta_{p}^{b,1}/\zeta_{3}^{b,1}$  &  $\zeta_{p}^{b,2}/\zeta_{3}^{b,2}$ &  $\zeta_{p}^{u}/\zeta_{3}^{u}$  &  $\zeta_{p}^{b,1}/\zeta_{3}^{b,1}$  &  $\zeta_{p}^{b,2}/\zeta_{3}^{b,2}$
 \\ \hline
$1$  &  $0.36 \pm 0.01$  &  $0.349 \pm 0.007$  &  $0.38 \pm 0.01$ & $0.36 \pm 0.01$  &  $0.348 \pm 0.005$  &  $0.37 \pm 0.01 $ & $0.355 \pm 0.007$  &  $0.349 \pm 0.005$  &  $0.38 \pm 0.01$
\\
$2$  &  $0.69 \pm 0.01$  &  $0.682 \pm .007$  &  $0.70 \pm .01$ &  $0.69 \pm 0.01$  &  $0.681 \pm .005$  &  $0.69 \pm .01$ &  $0.687 \pm 0.008$  &  $0.682 \pm .005$  &  $0.70 \pm .01$
\\
$3$  &  $1$  &  $1$  &  $1$ &  $1$  &  $1$  &  $1$ &  $1$  &  $1$  &  $1$
\\
$4$  &  $1.29 \pm 0.02$  &  $1.30 \pm 0.01$  &  $1.30 \pm 0.01$ &  $1.30 \pm 0.02$  &  $1.306 \pm 0.009$  &  $1.30 \pm 0.01$ &  $1.30 \pm 0.02$  &  $1.31 \pm 0.01$  &  $1.30 \pm 0.01$
\\
$5$  &  $1.57 \pm 0.06$  &  $1.60 \pm 0.03$  &  $1.61 \pm 0.03$ &  $1.58 \pm 0.04$  &  $1.60 \pm 0.02$  &  $1.61 \pm .03$ &  $1.58 \pm 0.04$  &  $1.60 \pm 0.03$  &  $1.60 \pm .03$
\\
$6$  &  $1.8 \pm 0.1$  &  $1.89 \pm 0.05$  &  $1.91 \pm 0.04$ &  $1.87 \pm 0.07$  &  $1.89 \pm 0.04$  &  $1.91 \pm 0.05$ &  $1.87 \pm 0.06$  &  $1.89 \pm 0.04$  &  $1.91 \pm 0.05$
\\
$7$  &  $2.1 \pm 0.1$  &  $2.18 \pm 0.08$  &  $2.22 \pm 0.06$ &  $2.1 \pm 0.1$  &  $2.18 \pm 0.05$  &  $2.22 \pm 0.07$ &  $2.15 \pm 0.08$  &  $2.18 \pm 0.07$  &  $2.22 \pm 0.06$
\\
$8$  &  $2.4 \pm 0.2$  &  $2.4 \pm 0.1$  &  $2.53 \pm 0.07$ &  $2.4 \pm 0.1$  &  $2.46 \pm 0.07$  &  $2.52 \pm 0.09$ &  $2.4 \pm 0.1$  &  $2.47 \pm 0.09$  &  $2.52 \pm 0.08$
\\
$9$  &  $2.7 \pm 0.2$  &  $2.7 \pm 0.1$  &  $2.84 \pm 0.09$ &  $2.7 \pm 0.2$  &  $2.74 \pm 0.09$  &  $2.8 \pm 0.1$ &  $2.7 \pm 0.1$  &  $2.8 \pm 0.1$  &  $2.8 \pm 0.1$
\\
$10$  &  $3.0 \pm 0.2$  &  $3.0 \pm 0.1$  &  $3.2 \pm 0.1$ &  $3.0 \pm 0.2$  &  $3.0 \pm 0.1$  &  $3.1 \pm 0.1$ &  $3.0 \pm 0.2$  &  $3.0 \pm 0.1$  &  $3.1 \pm 0.1$
\\ \hline
\end{tabular}
\end{center}
\caption{Multiscaling exponent ratios
$\zeta^u_p/\zeta^u_3$, $\zeta^{b,1}_p/\zeta^{b,1}_3$, and
$\zeta^{b,2}_p/\zeta^{b,2}_3$ with error bars for the order $p$ in the range $1\leq p \leq 10$ and runs R1, R3, and R4.}
\end{table*}

\begin{figure*}
\includegraphics[width=5.9cm,height=4cm]{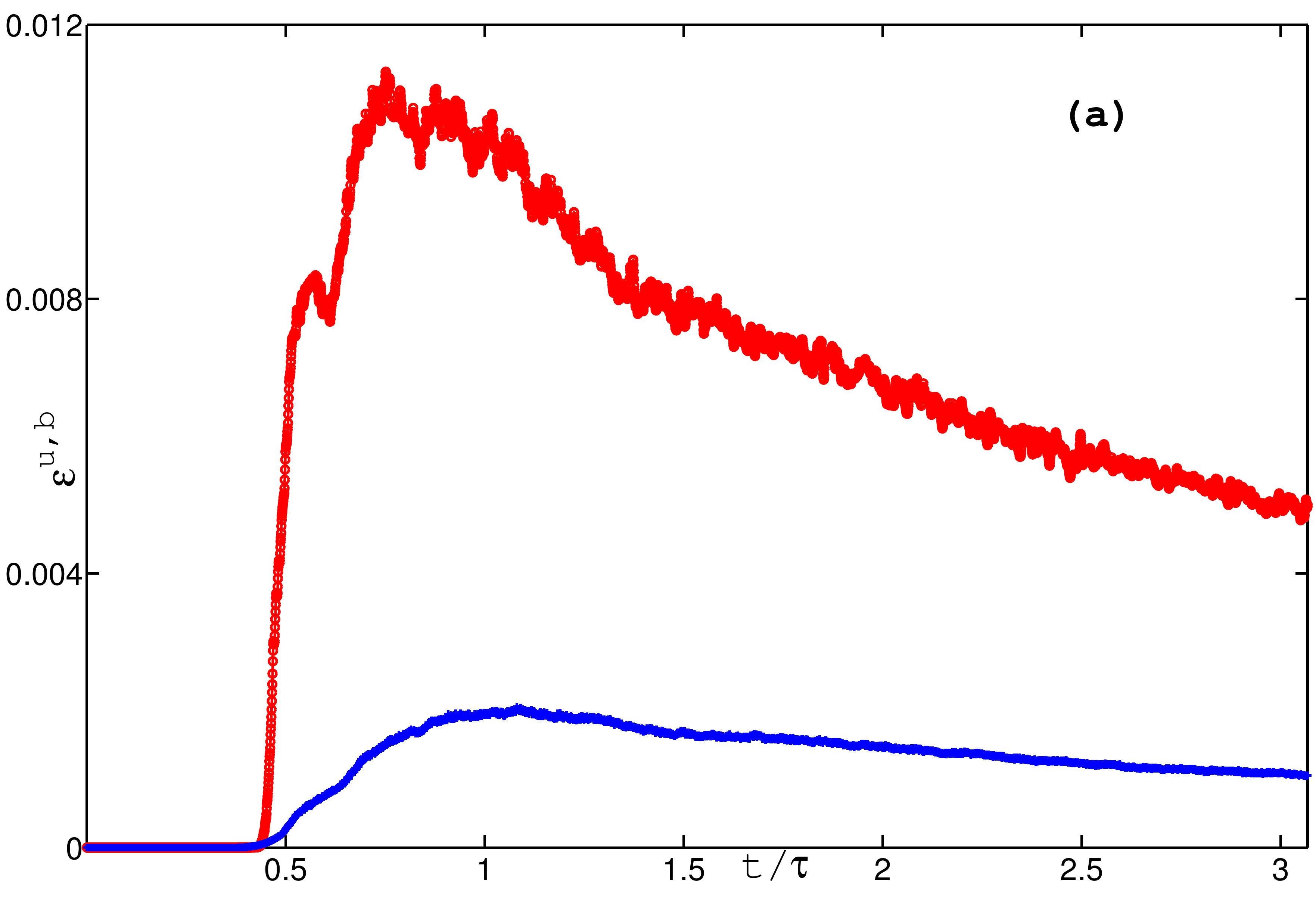}
\includegraphics[width=5.9cm,height=4cm]{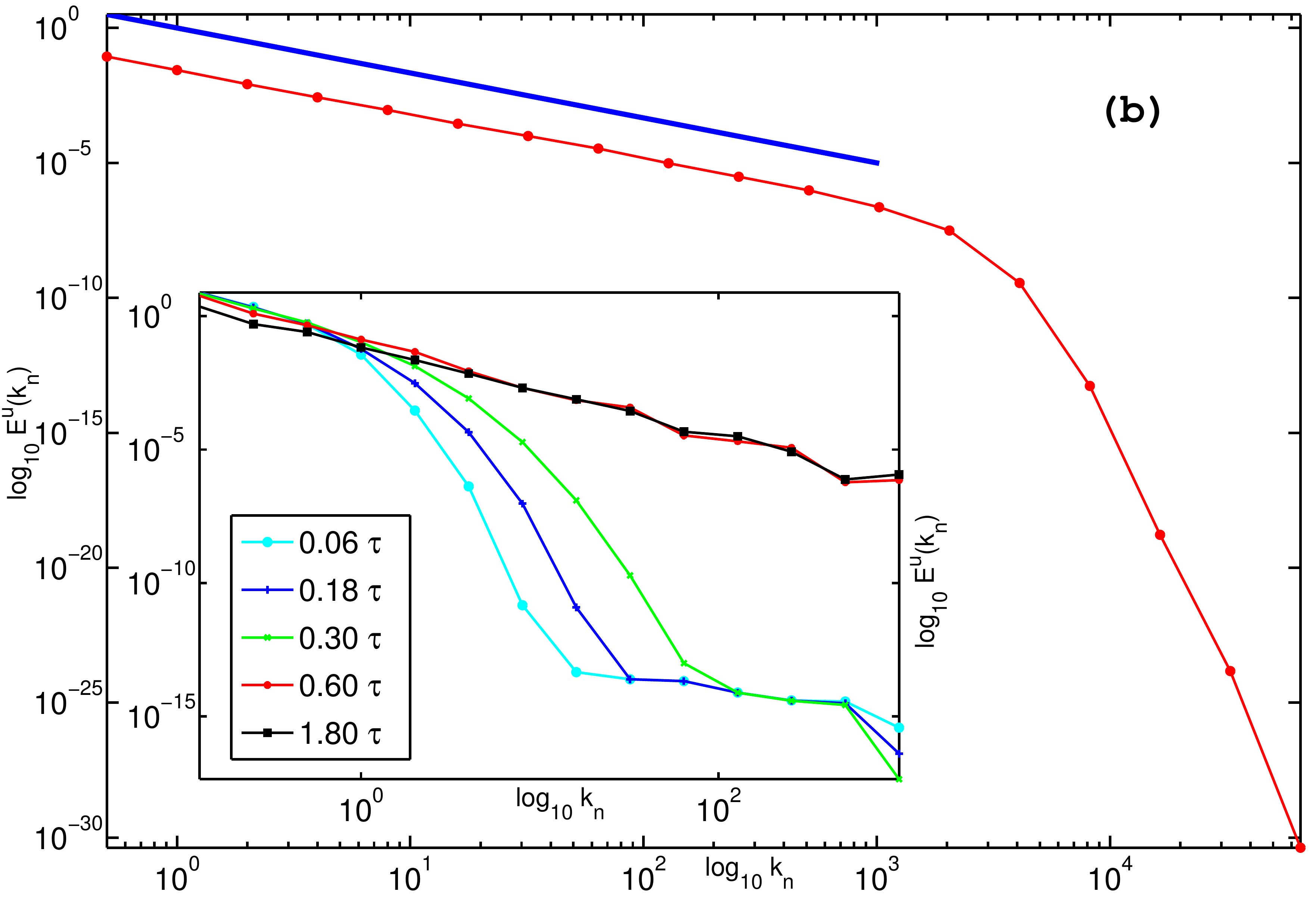}
\includegraphics[width=5.9cm,height=4cm]{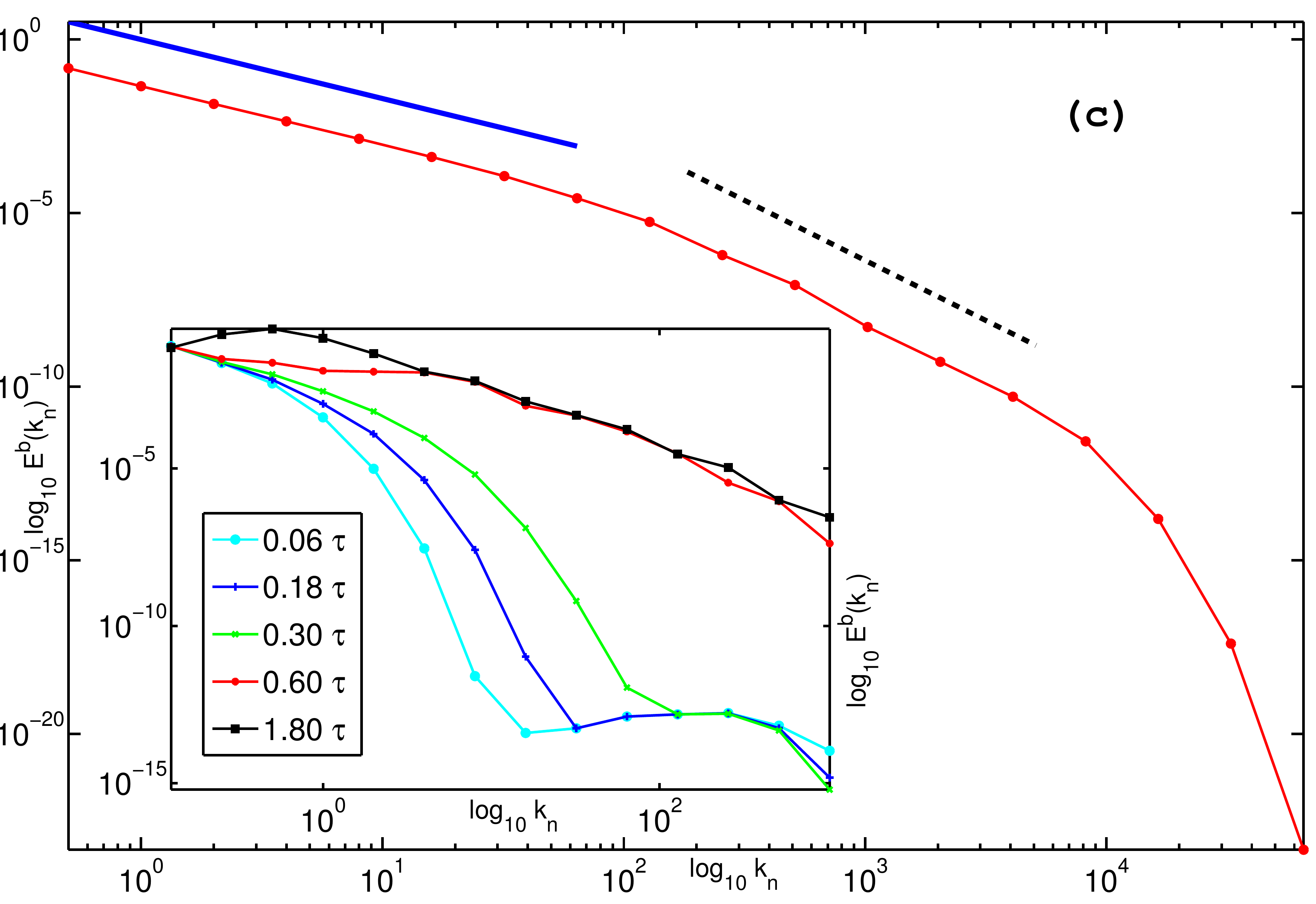}
\caption{(Color online) (a) Plots of  $\varepsilon^{u}$ (red, upper
curve) and  $\varepsilon^{b}$ (blue, lower curve) versus the rescaled time $t/\tau$.
(b) Plot of the kinetic-energy spectrum $E^u(k)$; the thick, blue line indicates
the K41-scaling power law. Inset : Time evolution of $E^u(k)$ during the initial period with
transients; the legend indicates the time at which each curve is obtained.
(c) Plot of the magnetic-energy spectrum $E^b(k)$; the thick, blue line
indicates K41 scaling in the low-$k$ region and the dashed, black line indicates
$k^{-3.5}$ scaling in the high-$k$ region . Inset : Time evolution of
$E^b(k)$ during the initial period with transients; the legend indicates
the time at which each curve is obtained. The data shown are for run R1.}
\label{figs:spectra1}
\end{figure*}

\begin{figure*}
\includegraphics[width=5.9cm,height=4cm]{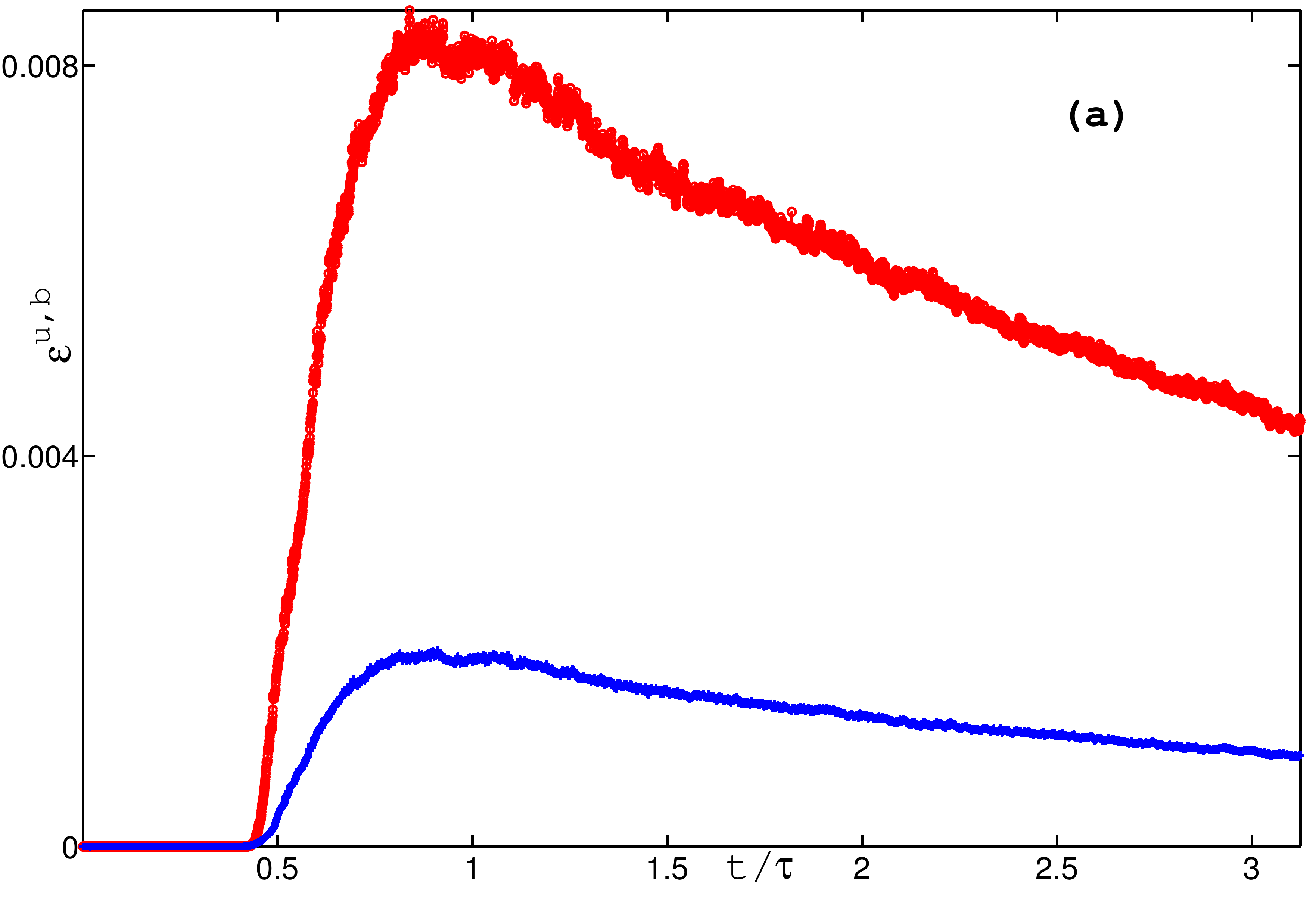}
\includegraphics[width=5.9cm,height=4cm]{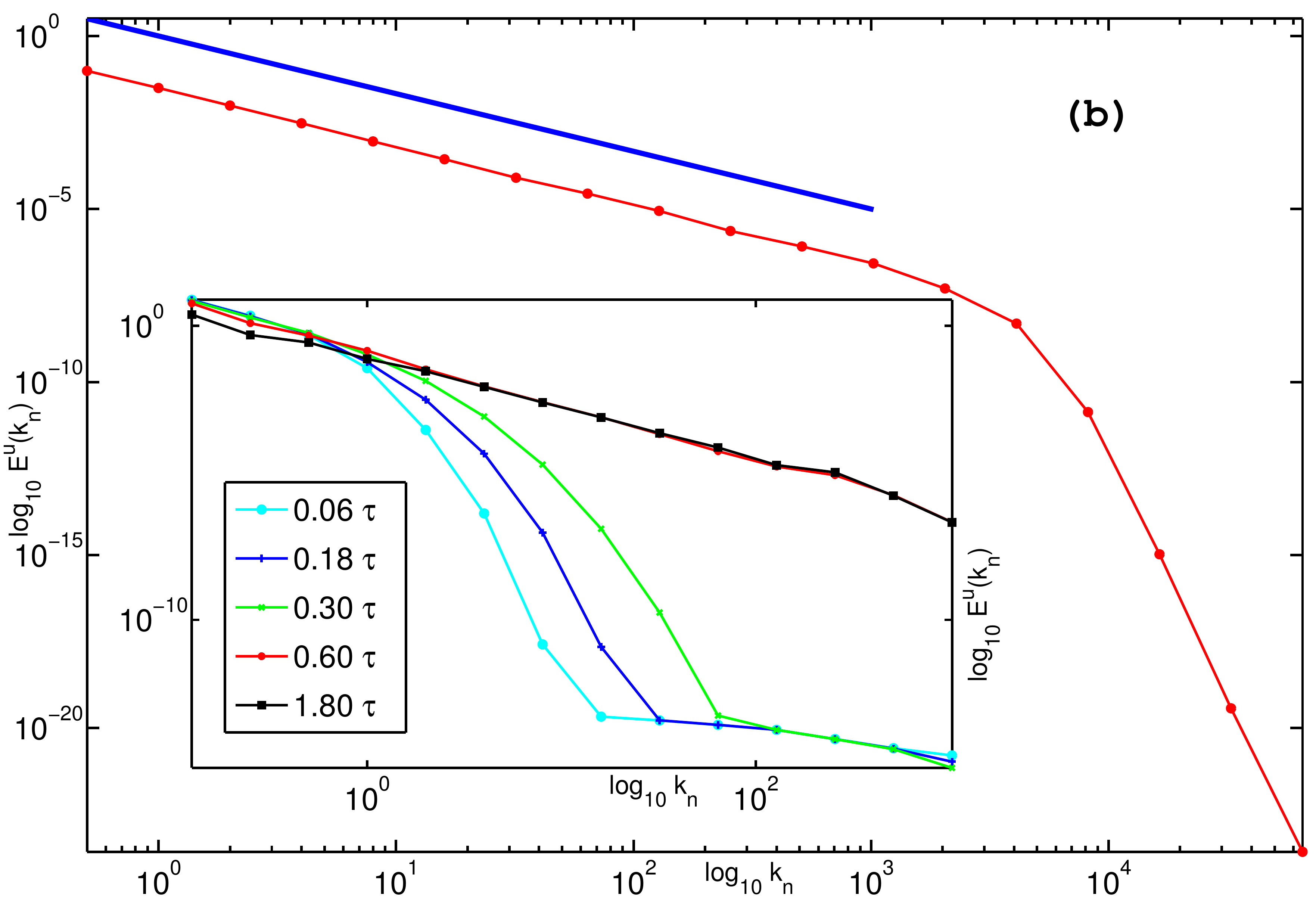}
\includegraphics[width=5.9cm,height=4cm]{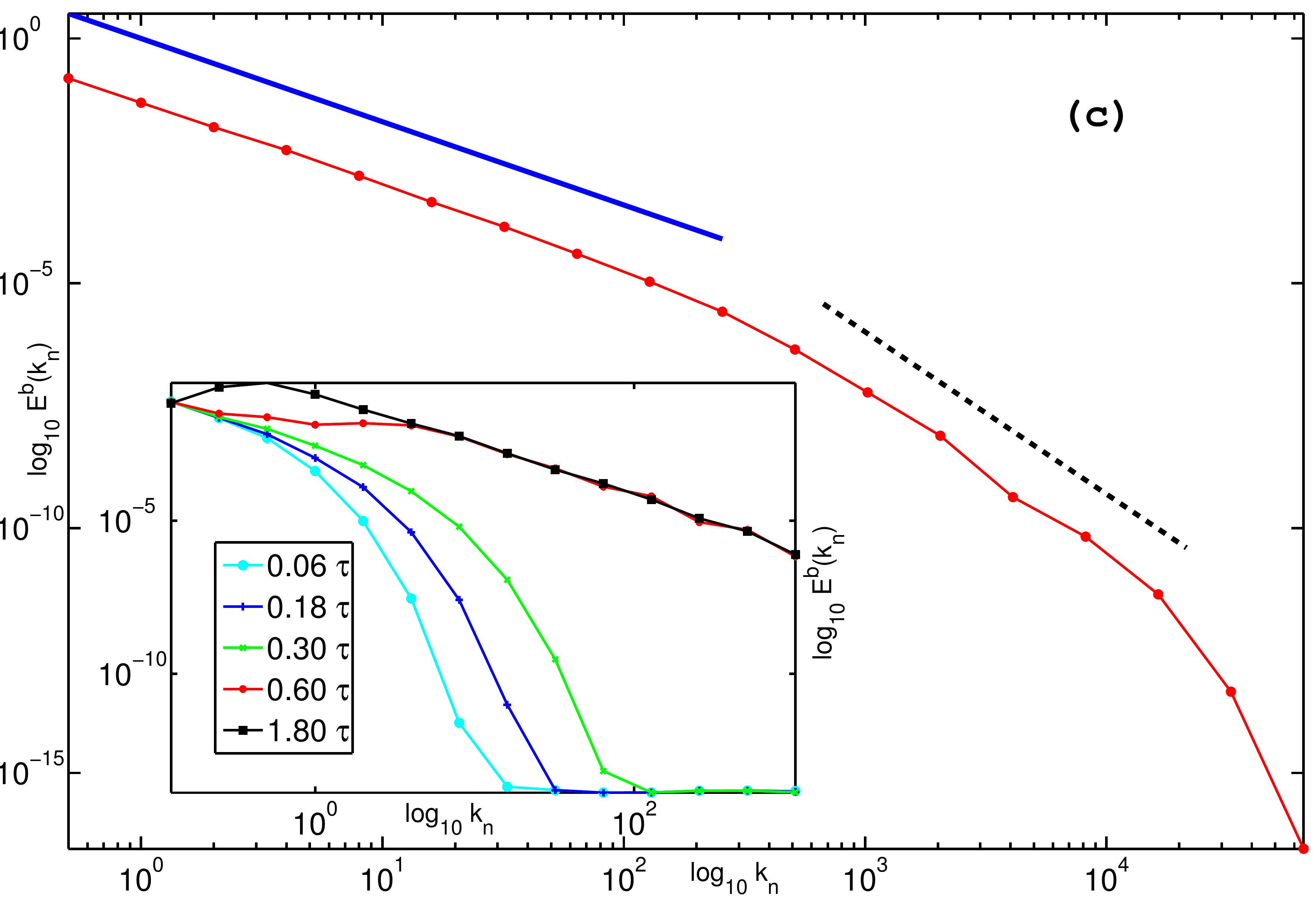}
\caption{(Color online) (a) Plots of  $\varepsilon^{u}$ (red, upper
curve) and  $\varepsilon^{b}$ (blue, lower curve) versus the rescaled time $t/\tau$.
(b) Plot of the kinetic-energy spectrum $E^u(k)$; the thick, blue line indicates
the K41-scaling power law. Inset : Time evolution of $E^u(k)$ during the initial period with
transients; the legend indicates the time at which each curve is obtained.
(c) Plot of the magnetic-energy spectrum $E^b(k)$; the thick, blue line
indicates K41 scaling in the low-$k$ region and the dashed, black line indicates
$k^{-3.5}$ scaling in the high-$k$ region . Inset : Time evolution of
$E^b(k)$ during the initial period with transients; the legend indicates
the time at which each curve is obtained. The data shown are for run R3.}
\label{figs:spectra3}
\end{figure*}

\begin{figure*}
\includegraphics[width=5.9cm,height=4cm]{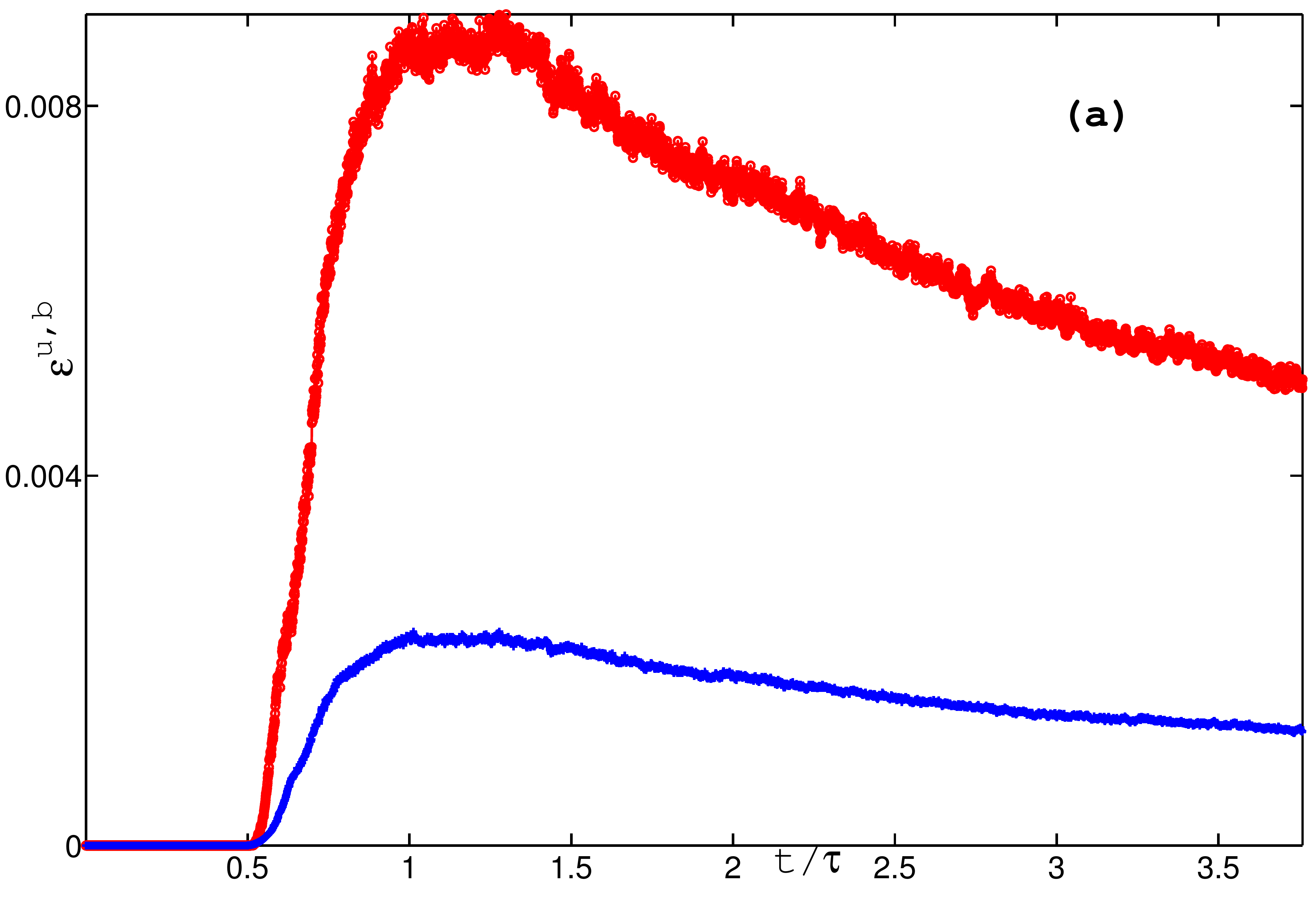}
\includegraphics[width=5.9cm,height=4cm]{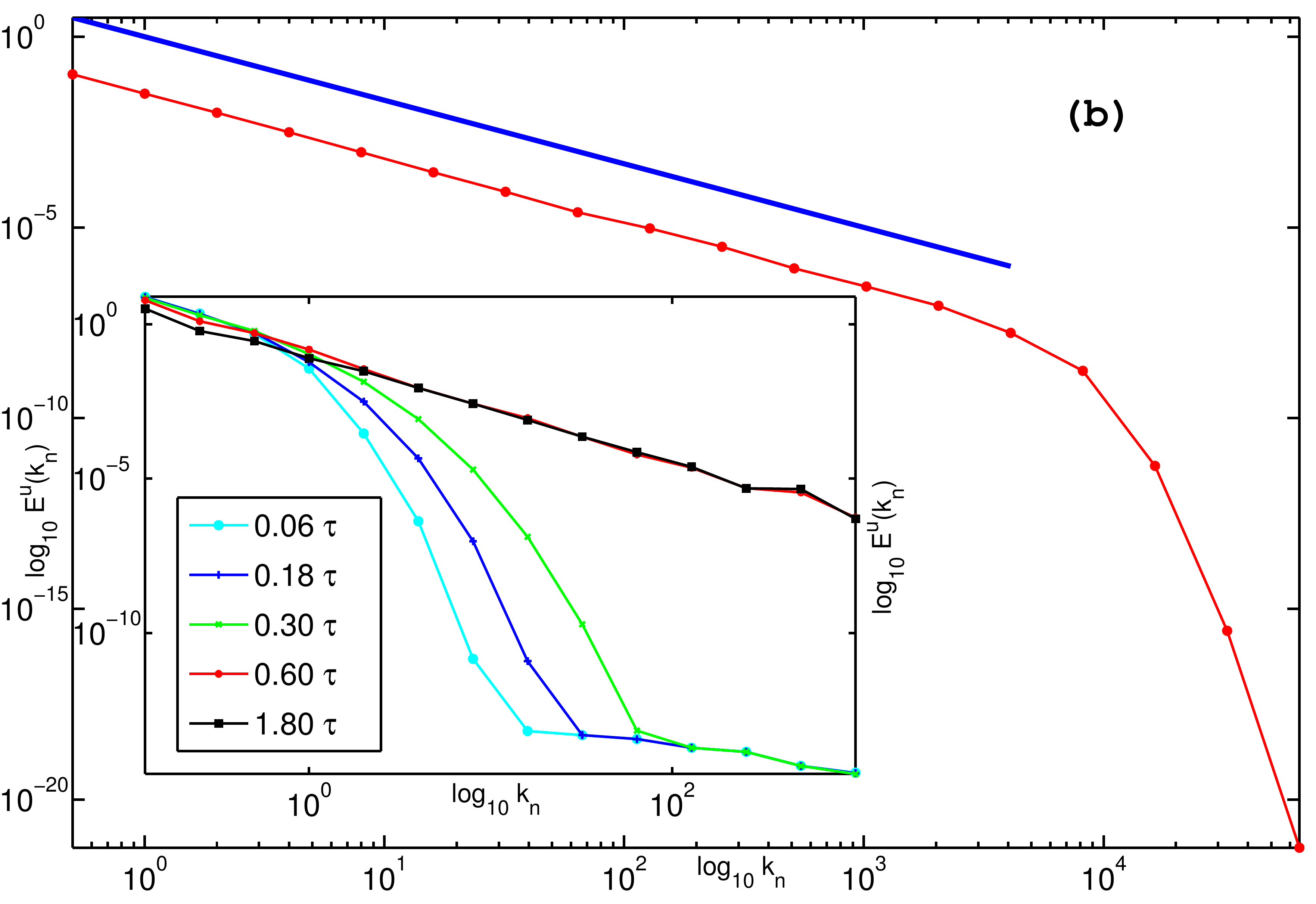}
\includegraphics[width=5.9cm,height=4cm]{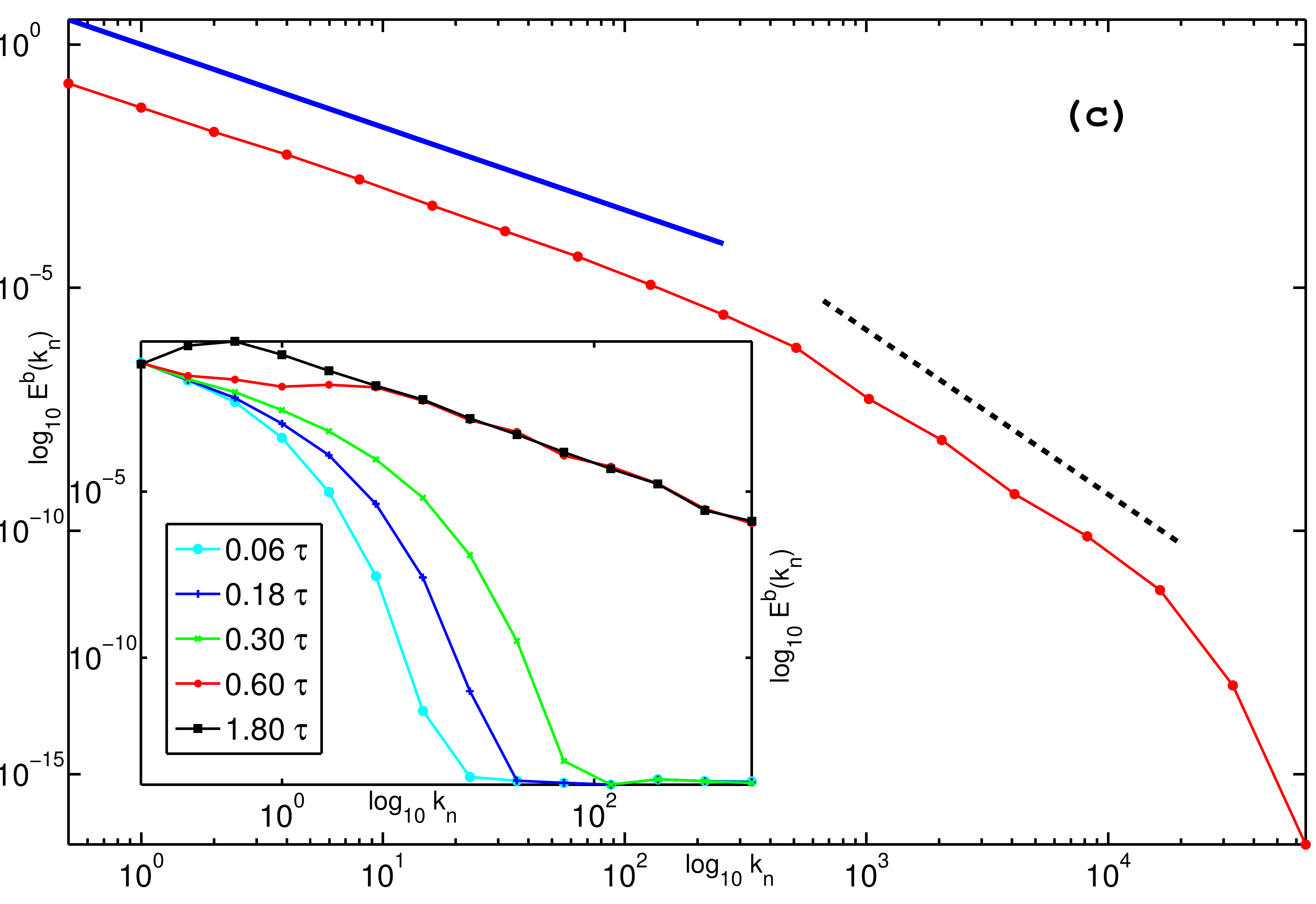}
\caption{(Color online) (a) Plots of  $\varepsilon^{u}$ (red, upper
curve) and  $\varepsilon^{b}$ (blue, lower curve) versus the rescaled time $t/\tau$.
(b) Plot of the kinetic-energy spectrum $E^u(k)$; the thick, blue line indicates
the K41-scaling power law. Inset : Time evolution of $E^u(k)$ during the initial period with
transients; the legend indicates the time at which each curve is obtained.
(c) Plot of the magnetic-energy spectrum $E^b(k)$; the thick, blue line
indicates K41 scaling in the low-$k$ region and the dashed, black line indicates
$k^{-3.5}$ scaling in the high-$k$ region . Inset : Time evolution of
$E^b(k)$ during the initial period with transients; the legend indicates
the time at which each curve is obtained. The data shown are for run R4.}
\label{figs:spectra4}
\end{figure*}


\begin{thebibliography}{}
\bibitem{foot1} We note, in passing, that $\zeta_p^u \simeq \zeta_p^{b,1} \simeq \zeta_p^{\rm fl}$, 
where $\zeta_p^{\rm fl}$ are the equal-time scaling exponents obtained for 
3D fluid turbulence~\cite{frisch95}.
\end{thebibliography}
\end{document}